\documentclass[acp, manuscript]{copernicus}

\usepackage{tikz}
\usetikzlibrary{patterns}
\usetikzlibrary{shapes}
\usetikzlibrary{decorations.text}
\usetikzlibrary{shapes.multipart}
\usetikzlibrary{arrows,shapes.geometric,positioning}
\usetikzlibrary{shapes,positioning,decorations.pathmorphing}
\usetikzlibrary{calc}
\usepackage{booktabs}
\usetikzlibrary{backgrounds}
\usepackage{varwidth}
\usetikzlibrary{fit}
\usepackage{bm,bbm}
\usepackage{amsmath,amsfonts,amssymb,mathrsfs}
\usepackage{siunitx}
\usepackage{latexsym,euscript,textcomp}
\usepackage{color,graphics,epsf,dcolumn}
\usepackage{epstopdf,ulem}
\usepackage{blindtext}
\usepackage{threeparttable}
\usepackage{url,hyperref,lineno,microtype,subcaption}
\usepackage{enumitem}

\allowdisplaybreaks

\begin{document}
\nolinenumbers

\title{Vegetation Impact on Atmospheric Moisture Transport under Increasing Land-Ocean Temperature Contrasts}

% \Author[affil]{given_name}{surname}

\Author[1,2,3]{Anastassia M.}{Makarieva}
\Author[1]{Andrei V.}{Nefiodov}
\Author[4]{Antonio Donato}{Nobre}
\Author[5,6,7]{Douglas}{Sheil}
\Author[8]{Paulo}{Nobre}
\Author[9]{Jan}{Pokorn{\'y}}
\Author[9]{Petra}{Hesslerov{\'a}}
\Author[3]{Bai-Lian}{Li}

\affil[1]{Theoretical Physics Division, Petersburg Nuclear Physics Institute, Gatchina  188300, St.~Petersburg, Russia}
\affil[2]{Institute for Advanced Study, Technical University of Munich, Lichtenbergstrasse 2 a, D-85748 Garching, Germany}
\affil[3]{USDA-China MOST Joint Research Center for AgroEcology and Sustainability, University of California, Riverside 92521-0124, USA}
\affil[4]{Centro de Ci\^{e}ncia do Sistema Terrestre INPE, S\~{a}o Jos\'{e} dos Campos, S\~{a}o Paulo  12227-010, Brazil}
\affil[5]{Forest Ecology and Forest Management Group, Wageningen University \& Research, PO Box 47, 6700 AA, Wageningen, The Netherlands}
\affil[6]{Center for International Forestry Research (CIFOR), Kota Bogor, Jawa Barat, 16115, Indonesia}
\affil[7]{Faculty of Environmental Sciences and Natural Resource Management, Norwegian University of Life Sciences, \AA s, Norway}
\affil[8]{Center for Weather Forecast and Climate Studies INPE, S\~{a}o Jos\'{e} dos Campos, S\~{a}o Paulo 12227-010, Brazil}
\affil[9]{ENKI, o.p.s., T{\v r}ebo{\v n}, Dukelsk{\'a} 145, Czech Republic}

%% The [] brackets identify the author with the corresponding affiliation. 1, 2, 3, etc. should be inserted.

\runningtitle{Vegetation Impact on Atmospheric Moisture Transport under Increasing Land-Ocean Temperature Contrasts}

\runningauthor{Makarieva et al.}

\correspondence{A.M.~Makarieva (ammakarieva@gmail.com)}

\received{}
\pubdiscuss{} %% only important for two-stage journals
\revised{}
\accepted{}
\published{}

%% These dates will be inserted by Copernicus Publications during the typesetting process.

\firstpage{1}

\maketitle

\begin{abstract}
Destabilization of the water cycle threatens human lives and livelihoods. Meanwhile our understanding of whether and how changes in vegetation cover could trigger abrupt transitions in  moisture regimes remains incomplete. This challenge calls for better evidence as well as for the theoretical concepts to describe it.  Here we briefly summarise the theoretical questions surrounding the role of vegetation cover in the dynamics of a moist atmosphere.  We discuss the previously unrecognized sensitivity of local wind power to condensation rate as revealed by our analysis of the continuity equation for a gas mixture. Using the framework of condensation-induced atmospheric dynamics, we then show that with the temperature contrast between land and ocean increasing up to a critical threshold,  ocean-to-land moisture transport reaches a tipping point where it can stop or even reverse. Land-ocean temperature contrasts are affected by both global and regional processes, in particular,  by the surface fluxes of sensible and latent heat that are strongly influenced by vegetation. Our results clarify how a disturbance of natural  vegetation cover, e.g., by deforestation, can disrupt large-scale atmospheric circulation and moisture transport. In view of the increasing pressure on natural ecosystems, successful strategies of mitigating climate change require taking into account the impact of vegetation on moist atmospheric dynamics. Our analysis provides a theoretical framework to assess this impact.  The available data for Eurasia indicate that the observed climatological land-ocean temperature contrasts are close to the threshold. This can explain the increasing fluctuations in the continental water cycle including droughts and floods and signifies a yet greater potential importance for large-scale forest conservation.
\end{abstract}

%%%%%%%%%%%%%%%%%%%%%%%%%%%%%%%%%%%%%%%%%%

\introduction 

Reliable water is crucial for human life. Long-term data indicate that in recent decades many regions of the world, including Eurasia and Western Europe, have been steadily losing soil moisture during the vegetative season \citep{gu19}. Other regions have experienced unprecedented droughts \citep{marengo15} and catastrophic floods have hit areas where they were previously unknown \citep{cornwall2021}.

Aridity stresses plants and modifies their influence on the water cycle and climate. These water-vegetation feedbacks present both risks and opportunities. The risks include the potential for aridity enhancement by vegetation dieback up to a complete switch from a wet to an arid climate state. The opportinities arise in the ability to slow down the aridification by preserving vegetation cover or to reverse the aridification and bring water back to otherwise arid and drought prone regions by human-mediated vegetation recovery.

Given the scale and nature of these trends and opportunities,  our understanding of the feedbacks remains inadequate. 
Theoretical studies of vegetation-water relations have long featured conceptual controversies. \citet{charney75} proposed that increased albedo from vegetation dieback should cool land, reduce the land-ocean temperature gradient, weaken ocean-to-land moisture advection and thus further enhance droughts. \citet{ripley76} objected that drying warms the land surface via a reduction in evapotranspiration and latent heat flux: a negative rather than a positive feedback. \citet{charney76}  replied that extra cooling over the drier land will manifest itself in the upper atmosphere and not on the surface, but agreed that the ultimate land-ocean temperature contrasts are model-dependent -- as was illustrated by later studies \citep{claussen97}.

\begin{figure*}[tb]
\begin{minipage}[p]{1\textwidth}
\centering
\includegraphics[width=0.9\textwidth,angle=0,clip]{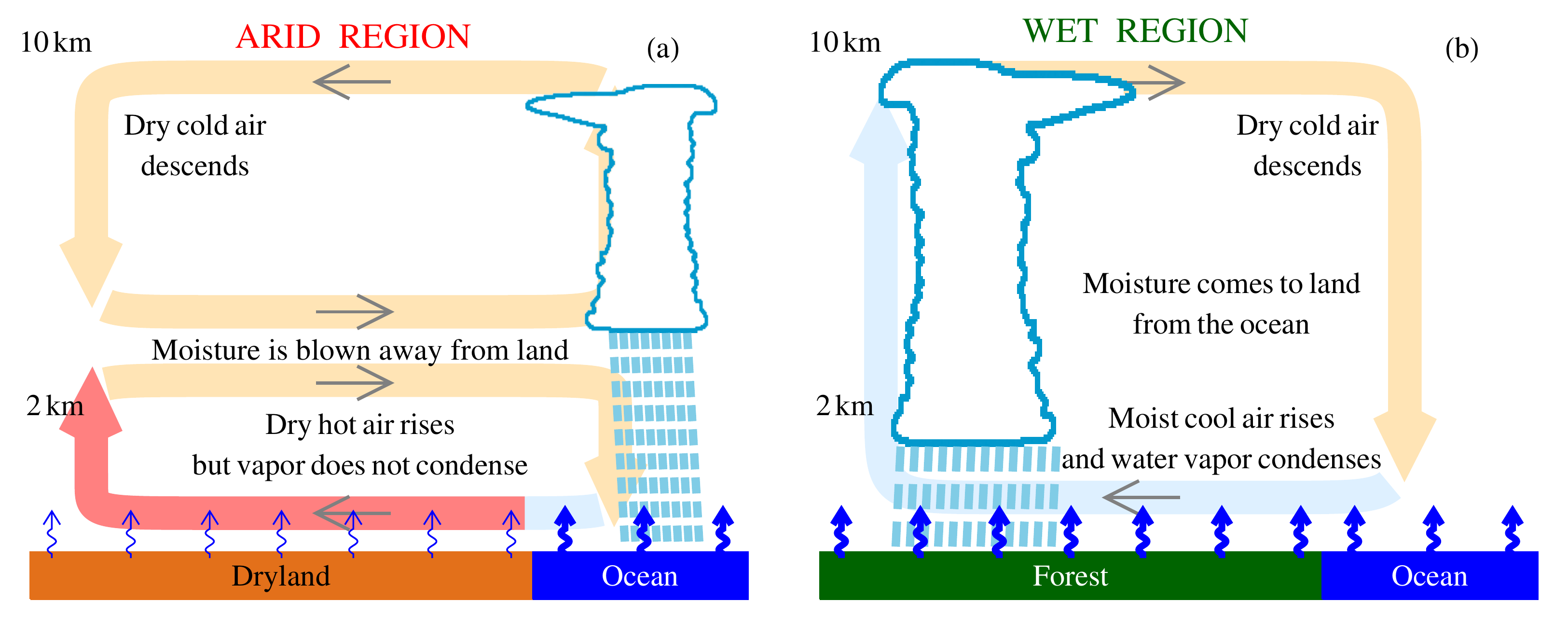}
\end{minipage}
\caption{
Air circulation between (\textbf{a}) the ocean and a hot and dry land \citep[cf. Fig. 3 of][]{charney75} and (\textbf{b}) the ocean and a cool and moist land. The ocean and the forest have higher evaporation rates (thick blue arrows) than the dryland (thin blue arrows).  The vertical profile of the pressure differences between the atmosphere over land and over the ocean in (a) and (b) can be qualitatively illustrated by, respectively, Fig.~2c and Fig.~2b of \citep{jcli15}. In (a), there is a pressure surplus over land in the lower troposphere at the height of the outflow, and pressure shortages above and below; in (b), there is a pressure shortage in the lower atmosphere and a pressure surplus in the upper atmosphere.}
\label{fig1}
\end{figure*}

This debate about evapotranspiration and latent heat persisted through many years. Discussions focused on
whether and how changes in vegetation could trigger an abrupt switch of ocean-to-land air circulation (Fig.~\ref{fig1}a,b).
\citet{Levermann2009} proposed that monsoonal regimes can switch via a tipping point involving a positive feedback of moisture advection on the land-ocean temperature contrast. The idea was that the more moisture comes from the colder ocean to condense over the warmer land, the more latent heat is released warming land even further. \citet{Boos2016a,Boos2016b} used a global climate model to demonstrate that such a scenario is physically implausible. To descend over the ocean, the air warmed by latent heat release over land must give the extra heat away (Fig.~\ref{fig1}b) -- otherwise the circulation would stop. The finite cooling rate limits the enhancement in circulation by latent heat release. In their reply, \citet{Levermann2016} did not specify any mechanism that could enable warm air to descend against buoyancy. The controversy persists: recently, \citet{Boers2017} proposed a similar concept of a drought-related tipping point but neglected previous discussions \citep{Levermann2009,Levermann2016,Boos2016a,Boos2016b}.

In a related context, \citet{Kuo2017} investigated the causality between deep convection and high tropospheric moisture content. Is high water vapor content a consequence of the atmosphere being moistened by convection, or, conversely, does high water vapor content trigger convection? \citet{Kuo2017}{\textquoteright}s  modelling results appeared to support the latter. In parallel, on a much wider spatial scale, another study similarly concluded that high transpiration by the Amazon rainforest during the late dry season moistens the atmosphere and triggers the beginning of the Amazon wet season well in advance of the arrival of the Intertropical Convergence Zone \citep{wright17}. Thus, the switch between circulation patterns in Fig.~\ref{fig1} can be enforced by atmospheric moistening via evapotranspiration.  \citet{pradhan19} also found that the onset of summer monsoon in Northeast India is preceded by enhanced transpiration. Recent studies indicate that evaporation during rainfall can be significant, highlighting the tight coupling between these processes \citep{murakami2021,jimenez2021}.

Theoretical debates in atmospheric sciences progress slowly. It took several years before \citet{Levermann2009}{\textquoteright}s  ideas were scrutinised.  Another example  pertains to the magnitude of the gravitational power of precipitation (how much wind power is spent to lift liqud water) in tropical cyclones.  Early studies concluded that the gravitational power of precipitation is a major term in the wind power budget of a tropical cyclone \citep{em88}. \citet{sabuwala15} reconsidered the issue and reached the same conclusion.  It was over 30 years since the original evaluation until  \citet{makarieva20,mpi4-jas} demonstrated, and \citet{er20} agreed, that in contrast to the previous conclusions this term is negligible.  Another slow debate concerns the role of dissipative heating in atmospheric circulations \citep[see][and references therein]{makarieva20,er20}.

A specific long-standing challenge in the analyses of vegetation-atmosphere feedbacks has been the inadequate representation of continental moisture convergence in global models \citep{ma06,hagemann11}. In the steady state, the net amount of atmospheric moisture brought to land by winds from the ocean (moisture convergence) must match the amount of liquid water that leaves land for the ocean as gravitational runoff. While runoff $\mathcal{R}$ is measured directly, moisture convergence $\mathcal{C}$ is model-derived. They do not generally match: instead of the equality, $\mathcal{R} = \mathcal{C}$, implied by mass conservation, the discrepancy between $\mathcal{C}$ and $\mathcal{R}$ can be of the order of   $100\%$ (as it is, for example, for the Amazon basin \citep{ma06}). For the continental moisture budget, $\mathcal{P} - \mathcal{E} = \mathcal{C}$, the underestimate of moisture convergence $\mathcal{C}$ implies that either precipitation $\mathcal{P}$ is underestimated, or  evapotranspiration $\mathcal{E}$ is overestimated, or both. Evapotranspiration is generally the least certain component of the terrestrial water cycle \citep[e.g.,][]{lugato2013}. Reliable analyses of vegetation-atmosphere feedbacks and their spatial and temporal propagation in large river basins such as the Amazon require a resolution to these inconsistencies \citep[e.g.,][]{salati91,zemp17b,zemp17a,molina19,ruisv20}.

Incomplete understanding of vegetation-water feedbacks has implications for models and resulting global climate projections. Recent studies demonstrate that  the warming that results from reduced transpiration more than compensates for the cooling that results from increased albedo, such that deforestation results in an elevation of local surface temperatures during the vegetative season by up to several kelvin \citep{hurina16,alkama16,hesslerova18}. The net change of global mean surface temperature resulting from changes of albedo and transpiration following a large-scale deforestation is estimated at about $\pm 0.05$~K (the sign varies among models) \citep{winckler19}. The gross changes, however, are more than an order of magnitude larger and comparable in magnitude to observed global warming. The nature of this fine balance between physically distinct effects has not been explained and requires an investigation. If it turns out to have resulted from model tuning, the impact of deforestation on climate destabilization may be greatly underestimated. The first systematic analyses of the forest control of cloud cover indicate that the previous assessments of the forest contribution to the maintenance of global surface temperature require a re-evaluation \citep{duveiller2021,cerasoli2021}.

This brief account demonstrates the many challenges and unresolved problems surrounding the field of moist atmospheric dynamics. The recent trajectory of environmental and climate research revolved more around the development of numerical models and empirical data gathering. Considering achievements to date, leading researchers have begun to re-emphasize the need for strong theoretical knowledge as a framing and foundation for effective climate science \citep{emanuel20}. Theory is required to judge and understand the  adequacy and outputs of numerical models. Vegetation-atmosphere feedbacks, with their complexity and profound implications for the humanity{\textquoteright}s well-being, appear to be the topic where new theoretical approaches could be particularly useful. 

Here we provide some insights from {\it biotic pump} \citep{hess07} and the associated condensation-induced atmospheric dynamics (CIAD).  These concepts were invoked to explain spatial and temporal precipitation patterns in various regions \citep[e.g.,][]{andrich13,poveda14,molina19}. These triggered multiple discussions \citep{meesters09,makarievagvg09,angelini11,mgl13,JMR1,JMR2,makarieva19,pearce20}. The main implication of the biotic pump for the vegetation-atmosphere dynamics is that large-scale forests, by  generating and maintaining atmospheric moisture through transpiration, can power the ocean-to-land  winds and the associated atmospheric moisture transport. Essentially, water vapor removal from the gas phase produces non-equilibrium pressure gradients that generate both vertical and horizontal wind.

In Section~\ref{cwp} we briefly introduce the main equations of CIAD with an emphasis on how local wind power can be estimated from the continuity equation. We highlight and explain the sensitivity of the wind power to the formulation of the condensation rate. Next in Section~\ref{chtg} we formulate CIAD in an integral form that allows the estimation of the role of the horizontal temperature differences for the moisture transport. In Section~\ref{exp}  we use climatological data to estimate the relevant quantities for Eurasian regions. We discuss how the regional cooling provided by the transpiring vegetation cover can buffer the land-ocean temperature differences and prevent the drought-related tipping points in a climate where land is warming faster than the ocean. In the concluding section we discuss the importance of fostering constructive debates of theoretical problems related  to the biotic pump concept and outline a few implications of the obtained results for current climate policies.

\section{Condensation rate and wind power}
\label{cwp}

Two fundamental vertical scales characterize the moist atmosphere of the Earth. One is the hydrostatic height $h \equiv -p/(\partial p/\partial z) =  RT/Mg \sim 10$~km determined by the interplay between  the gravitational and internal energy of the atmospheric gases. Another is the vertical scale height for the condensable gas, water vapor, $h_c \equiv -p_v/(\partial p_v/\partial z) = RT^2/L\Gamma \sim 2$~km,  that is determined by the interplay between the cooling rate of ascending air and latent heat release during any resulting condensation. Here $p$ is air pressure, $p_v$ is partial pressure of saturated water vapor, $R = 8.3$~J~mol$^{-1}$~K$^{-1}$ is the universal gas constant, $T$ is absolute temperature, $M \simeq 29$~g~mol$^{-1}$ is mean molar mass of atmospheric gases, $g$ is the acceleration of gravity, $L \simeq 45$~kJ~mol$^{-1}$ is the latent heat of vaporization, $\Gamma \equiv - \partial T/\partial z\simeq 6.5$~K~km$^{-1}$ is the vertical lapse rate of air temperature.

The fact that $h_c \ll h$ means that the vertical gradient of water vapor partial pressure is strongly non-equilibrium. This allows the formulation of the rate of potential energy release (W m$^{-3}$) during condensation in the ascending air as
\begin{equation}\label{Pi}
s \equiv - w p_v \left(\frac{1}{h_c} - \frac{1}{h}\right) \equiv w p   \frac{\partial \gamma}{\partial z} \equiv - w f_e  , 
\quad f_e \equiv  -\frac{\partial p_v}{\partial z} + \frac{p_v}{p}  \frac{\partial p}{\partial z} \equiv  \frac{p_v}{h_{\gamma}},
\end{equation}
where $h_{\gamma} \equiv - \gamma / (\partial \gamma/\partial z) = 1/ (h_c^{-1} - h^{-1})$,  $\gamma \equiv p_v/p$, $w$ is the vertical air velocity and $f_e$ has the meaning of a vertical force associated with the non-equilibrium partial pressure  gradient of water vapor. 

The main proposition of the biotic pump concept -- and the underlying condensation-induced atmospheric dynamics  -- is 
a power source for atmospheric circulation \citep{hess07,mg10,makarieva19}. Applied locally in a hydrostatic horizontally isothermal saturated atmosphere, where all wind power is generated by horizontal pressure gradients, this proposition takes the form
\begin{equation}
\label{K}
u \frac{\partial p}{\partial x} = s ,
\end{equation}
where $u$ is horizontal air velocity directed along  $x$-axis.

Theoretical relation (\ref{K}) agreed with observations in different atmospheric contexts, including general atmospheric circulation, the Amazon basin  and the more compact circulation patterns like hurricanes and tornadoes (see \citep{makarieva19} and references therein). In these compact vortices partial pressure of water vapor $p_v$ sets the scale for maximum wind velocity $u_{\rm max} = \sqrt{2p_v /\rho} \sim 70$~m~s$^{-1}$, where $\rho \simeq 1$~kg~m$^{-3}$ is air density. This generality -- i.e., the validity of Eq. (\ref{K}) across several orders of magnitude for vertical velocity $w$ --  is satisfying for a theorist and incentivizes efforts to understand the underlying mechanisms and their implications more comprehensively.

Noting the different equivalent expressions for $s$ \eqref{Pi}, we observe similarity between $s$ and the term containing vertical velocity in the continuity equation expressed in terms of pressure:
\begin{equation}\label{contp}
w \left(\frac{\partial p_v}{\partial z} - \frac{p_v}{p_d}  \frac{\partial p_d}{\partial z} \right) + u \left(\frac{\partial p_v}{\partial x} - \frac{p_v}{p_d}  \frac{\partial p_d}{\partial x} \right) =\sigma .
\end{equation}
Here $p_d = p - p_v$ is the partial pressure of dry air, $\sigma = \mathcal{S}RT$  is the rate of phase transitions in power units (W~m$^{-3}$),  $\mathcal{S}$ (mol~s$^{-1}$~m$^{-3}$) is the molar rate of phase transitions (see Eqs.~(1), (6) and (8) of \citep{g12} and Eq.~\eqref{eq10} in the Appendix). Such a representation of the continuity equation is only possible for an ideal gas with its equation of state relating molar density and pressure.

The relation between $s$ (\ref{Pi}) and 
\begin{equation}\label{defz}
s_d \equiv - w p_v \left(\frac{1}{h_c} - \frac{1}{h_d}\right) \equiv w \left(\frac{\partial p_v}{\partial z} - \frac{p_v}{p_d}  \frac{\partial p_d}{\partial z} \right) \equiv  w p_d \frac{\partial \gamma_d}{\partial z},
\end{equation}
where $\gamma_d \equiv p_v/p_d$,  $h_d \equiv RT/M_d g$, $M_d$ is the molar mass of dry air, is 
\begin{equation}\label{ccz}
s \equiv (1 - \gamma) s_d. 
\end{equation}
They differ by a small magnitude $\gamma \equiv p_v/p \ll 1$.

At constant relative humidity, $p_v$ grows with increasing temperature $T$ in accordance with the Clausius-Clapeyron equation
\begin{equation}\label{CC}
\frac{dp_v}{p_v} = \xi \frac{dT}{T}, \quad \xi \equiv \frac{L}{RT}.
\end{equation}

Assuming relative humidity to be constant and the air to be isothermal in the horizontal plane\footnote{While turbulent diffusion is not explictly accounted for in the continuity equation, it is implicitly present in the condition $\partial p_v/\partial x = 0$, i.e., turbulent diffusion is what ensures constant relative humidity on an isothermal plane.}, such that $\partial p_v/\partial x = 0$, we can write the continuity equation~(\ref{contp}) in the following form:
\begin{equation}\label{eq8*}
u \frac{\partial p}{\partial  x}= u \frac{\partial p_d}{\partial  x}= \frac{1}{\gamma_d}\left(s_d - \sigma \right) . 
\end{equation}

Since the vertical motions associated with adiabatic cooling are the most important mechanisms that bring moist air to saturation,  
one can argue that condensation rate can be approximated as
\begin{equation}
\label{CJ}
\sigma \equiv \alpha s_d, 
\end{equation}
where $\alpha \lesssim 1$ \citep[see, e.g.,][]{JMR2}. Then Eq.~(\ref{eq8*}) can be re-written as

\begin{equation}
\label{cJ}
u\frac{\partial p}{\partial x} = \frac{s_d}{\gamma_d}(1-\alpha) , 
\end{equation}
where $\gamma_d \equiv p_v/p_d$. Equation (\ref{cJ}) contains a product of a large factor $\gamma_d^{-1} \sim 10^2$ and an unknown small factor $1 -\alpha \ll 1$.  Thus, assuming that a certain (a priori unknown) $\alpha \simeq 1$ matches the observations, a mere $10\%$ reduction of $\alpha$, while still obeying $\alpha \simeq 1$, would lead to an order of magnitude overestimate  of $u \partial p/\partial x$. Conversely, any $p_d/p < \alpha < 1$ will produce unrealistically low (down to zero) values of $u \partial p/\partial x$.

To illustrate this, putting $\alpha = p_d/p + \Delta \alpha$ into Eq.~(\ref{cJ})  and taking into account Eq.~\ref{ccz}, we obtain
\begin{equation}
\label{cJ2}
u\frac{\partial p}{\partial x} = \frac{s_d }{\gamma_d}\left(1- \frac{p_d}{p}- \Delta \alpha \right) =  s \left(1 -  \frac{p}{p_v} \Delta \alpha \right).
\end{equation}
In the atmosphere of Earth with a typical value of $p_v \sim 20$~hPa, we have $p/p_v = 50$ and $p_d/p = 0.98$.  With an exemplary $\Delta \alpha = 0.018$, we have $\alpha = 0.998$.  Then the term in braces in the right-hand side of Eq.~(\ref{cJ2}) is equal to $0.1$ and we obtain a wind power ten times less than the observed.   Conversely, with $\Delta \alpha  = -0.2$ and $\alpha = 0.78$, the term in braces is equal to $11$ and we obtain an order of magnitude higher wind power. In both cases, $\alpha = 0.998$ and $\alpha = 0.78$ the specification $\alpha < 1$ and are sufficiently close to unity to satisfy the stipulation that the vertical motions and gradient make a dominant contribution to condensation rate (\ref{CJ}).   Nonetheless, the derived wind powers differ greatly -- the $20\%$ change in condensation rate $\sigma$ relative to $s_d$ (\ref{CJ}) has caused wind power to vary by two orders of magnitude.  While it follows from Eq.~(\ref{K}), that $\alpha \lesssim 1$ in the continuity equation~(\ref{cJ2}), the main dynamic equation of CIAD, Eq.~(\ref{K}), cannot be derived, even approximately, from the parameterization $\alpha \lesssim 1$ \citep[cf.][]{JMR2}.

\begin{table}[!ht]
    %\normalsize        
    \caption{Physical meaning of the two expressions for condensation rate.}\label{tab}
    \begin{threeparttable}
    \centering    
    \begin{tabular}{llll}
    \hline \hline 
   &&& \\
    \multicolumn{1}{c}{Equilibrium} & \multicolumn{1}{c}{Condensation rate $\sigma$} & \multicolumn{1}{c}{{\textquotedblleft}Horizontal{\textquotedblright} power} &\multicolumn{1}{c}{{\textquotedblleft}Vertical{\textquotedblright} power} \\ [2ex]
    \hline
 $\dfrac{\partial p}{\partial z} + \rho g = 0$ & $s \equiv wp \dfrac{\partial \gamma}{\partial z}$ & $ u \dfrac{\partial p}{\partial x} = s$ & $w\left(\dfrac{\partial p}{\partial z} + \rho g \right) = 0$ \\ [2ex]
 $\dfrac{\partial p_d}{\partial z} + \rho_d g = 0$ & $s_d \equiv wp_d \dfrac{\partial \gamma_d}{\partial z}$ & $u \dfrac{\partial p}{\partial x} = 0$ & $ w\left(\dfrac{\partial p}{\partial z} + \rho g \right) = s_d + w p_v\dfrac{\epsilon }{h_d}$ \\    [2ex]
\hline   \hline
    \end{tabular}
\begin{tablenotes}[para,flushleft]
Here $\rho = MN$ and $\rho_d = M_d N_d$ are the densities of total air and dry air, respectively; $\epsilon \equiv M_v/M_d -1$. {\it If} one sets condensation rate as indicated, then the expression for the {\textquotedblleft}horizontal{\textquotedblright} wind power follows from the continuity equation and $\partial p_v/\partial x = 0$. Assuming {\it additionally} the equilibrium condition yields the expression for the {\textquotedblleft}vertical{\textquotedblright} wind power.
\end{tablenotes}
\end{threeparttable}
\end{table}

We note that condensation-induced power $s$ (\ref{K}) was introduced from basic principles without referring to the continuity equations \eqref{eq67}. Thus, its relation to condensation rate $\sigma$ (\ref{eq10}) is not {\it a priori} obvious. Using Eq.~(\ref{K}) to replace  $u\partial p/\partial x$ with $s$ in Eq.~(\ref{eq8*}) gives $s = \sigma$, such that
\begin{equation}\label{cp}
u\frac{\partial p}{\partial x} = \sigma .
\end{equation}
While the estimate of wind power from the continuity equation is extremely sensitive to the formulation of the {\it a priori} unknown condensation rate  $\sigma$, the realistic values of wind power are consistent with the continuity equation under the assumption that wind power is exactly equal to condensation rate. This is an independent theoretical argument in favor of CIAD.

In the CIAD framework, the high sensitivity of wind power to the magnitude of $\sigma$ can be interpreted as follows. Let us consider the first row in Table~\ref{tab}. If condensation rate is put equal to $s$ (\ref{Pi}), then it follows from the continuity equation (\ref{contp}) that the wind power generated by the {\it horizontal} pressure gradient is equal to condensation rate. The physical justification for the expression for $s$ (\ref{Pi}) consists in the idea that the gradient of the partial pressure of water vapor is non-equilibrium {\it relative to the hydrostatic equilibrium of air as a whole}. When the air as a whole is in hydrostatic equilibrium, the wind power generated by the vertical pressure gradient is zero (work of the upward pressure gradient force is compensated by gravity).

Now let us consider the second row in Table~\ref{tab}. As compared to the first row, in the expression for condensation rate total pressure $p$ is replaced by partial pressure $p_d$ of dry air, $\sigma = s_d$. In this case it follows from the continuity equation that the wind power generated by the horizontal pressure gradient is zero. Indeed, putting $\alpha = 1$ in Eq.~(\ref{cJ}) (or, equivalently, $\Delta \alpha = \gamma$ in Eq.~(\ref{cJ2})), gives $u\partial p/\partial x = 0$.

If we assume that now the dry air is in hydrostatic equilibrium (which would justify using $p_d$ instead of $p$ in the expression for $\sigma$), we find that now the wind power generated by the {\it vertical} pressure gradient is equal to condensation rate -- plus an additional term proportional to the difference in the molar masses of the water vapor and dry air. This term is relatively small and its physical nature is not related to condensation\footnote{This term represents an additional work by gravity associated with the fact that it is the lighter gas (water vapor) that is compressed in the vertical relative to equilibrium. With $h_c \simeq 2$~km, $h_d \simeq h \simeq 10$~km and $M_v/M_d \simeq 0.6$ this term increases the absolute magnitude of the wind power by approximately 10\%.}.
In an atmosphere where $M_v = M_d = M$ the symmetry would be exact: if the {\textquotedblleft}horizontal{\textquotedblright} wind power is equal to condensation rate, then the {\textquotedblleft}vertical{\textquotedblright} one is zero, and vice versa.

As condensation rate changes from $s$ to $s_d$ \eqref{defz},  the wind power generated by the horizontal pressure gradient diminishes from $s$ to zero, while the wind power generated by the vertical pressure gradient grows from zero to, approximately,  $s_d$.
(The atmosphere changes then from a hydrostatic to a non-hydrostatic with the non-equilibrium vertical pressure
difference of the order of $p_v$.)  At intermediate values the power of condensation is allocated to both vertical and horizontal dimensions.
These considerations indicate that equations (\ref{K}) and (\ref{cp}) should remain valid for describing condensation-induced
circulation patterns if the kinetic energy of the vertical motion is much less than the kinetic energy of the horizontal motion. For example, it remains valid even in tornadoes where the air is non-hydrostatic, but the squared vertical velocity is still a few times less than the squared horizontal velocity \citep[e.g.,][Fig.~1B]{tor11}.

\section{Wind power and horizontal temperature gradient}
\label{chtg}

Applying Eq.~(\ref{cp}) to the non-isothermal case, i.e., solving it together with the continuity equation in the form of Eq.~(\ref{eq8*}), leads to the following modification of Eq.~(\ref{K}) (see \citep{mgn14} and Appendix A for derivation details)
\begin{equation}\label{tg}
u \frac{\partial p}{\partial x} =  s + u \frac{\partial p_v}{\partial x} = wp \frac{\partial \gamma}{\partial z} + 
u \frac{\partial p_v}{\partial x}.
\end{equation}
This shows that if the partial pressure of water vapor grows along the horizontal air streamline,
the condensation-induced wind power is diminished. While condensation reduces air pressure, evaporation 
adds gas to the flow and thus increases air pressure along the streamline inhibiting the 
condensation-induced air flow.

Using the Clausius-Clapeyron equation~(\ref{CC}) we can re-write Eq.~(\ref{tg}) as follows \citep{mg10,mgn14}:
\begin{equation}\label{horgr}
-\frac{\partial p}{\partial x} =\frac{w}{u} \frac{p_v}{h_\gamma} -  \frac{\partial p_v}{\partial x}, \quad  
 \frac{\partial p_v}{\partial x} = p_v  \frac{\xi}{T} \frac{\partial T}{\partial x} .
\end{equation}
Linearizing Eq.~(\ref{horgr}) by assuming that $w/u \sim h_w/l$, where $h_w$ and $l$ are the characteristic vertical and horizontal scales of the moisture inflow in the lower atmosphere, and $\partial p/\partial x \sim \Delta p/l$, $\partial p_v/\partial x \sim \Delta p_v/l$, $\partial T/\partial x \sim \Delta T/l$, we obtain
\begin{equation}\label{beta}
-\Delta p = p_{v\mathrm A}  \frac{h_w}{h_\gamma} - p_{v\mathrm D}\xi \frac{\Delta T}{T}  = p_{v\mathrm D} \left[\beta - \xi \frac{\Delta T}{T}(1 - \beta)\right],
\end{equation}
where $ \Delta p \equiv p_{\mathrm A} - p_{\mathrm D}$ and $ \Delta p_v \equiv p_{v\mathrm A} - p_{v\mathrm D}= p_{v\mathrm D}\xi \Delta T/T$. The quantities of pressure with indices ${\mathrm D}$ and ${\mathrm A}$ refer to the region that exports moisture (the {\textquotedblleft}donor{\textquotedblright}) and the region that receives this moisture (the  {\textquotedblleft}acceptor{\textquotedblright}), respectively. Factor $\beta \equiv h_w/h_\gamma \lesssim 1$ corresponds to $1 -\zeta$ of \citep{jas13} and describes the completeness of condensation in the ascending air, i.e., the share of water vapor that has condensed by the altitude when the air flow changes its horizontal direction (e.g., for the schematic circulation patterns in Figs.~\ref{fig1}a and \ref{fig1}b we have, respectively, $h_w = 2$~km and $10$~km).

Equation~(\ref{beta}) shows that, for a given $\beta < 1$, when temperature increases significantly along the horizontal air flow, the negative pressure difference $\Delta p < 0$ that drives the flow diminishes and, at sufficiently large $\Delta T$, can become zero. In this situation, condensation in the ascending air removes as much water vapor as is added to the horizontally moving air near the surface. As a result, the warmer area is locked for condensation-induced air circulation and the condensation-induced moisture inflow ceases. Equation~(\ref{beta}) is a manifestation of the general principle that if condensation and evaporation are {\it not} spatially separated (e.g., if evaporation in the acceptor region is compensated by condensation), no condensation-induced circulation can develop.

When condensation is complete ($\beta = 1$), all the additional water vapor that evaporates into the air as it moves from the donor to acceptor region ultimately condenses in the acceptor  region. The temperature gradient makes no impact on $\Delta p$.

Under global change, land surface is warming faster than the ocean due to its lower heat capacity and deforestation that reduces transpiration and elevates surface temperatures during the warmer season \citep[e.g.,][]{alkama16}. Thus, the temperature differences between land (that receives moisture from the ocean) and the ocean (which supplies moisture to land) can be expected to grow. It is thus important to estimate observed $\Delta T$ values to find out whether major ocean-to-land moisture flows may be close to a tipping point (when the term in square brackets  in Eq.~(\ref{beta}) becomes zero).

\begin{figure*}[tb]
\begin{minipage}[p]{1\textwidth}
\centering
\includegraphics[width=0.8\textwidth,angle=0,clip]{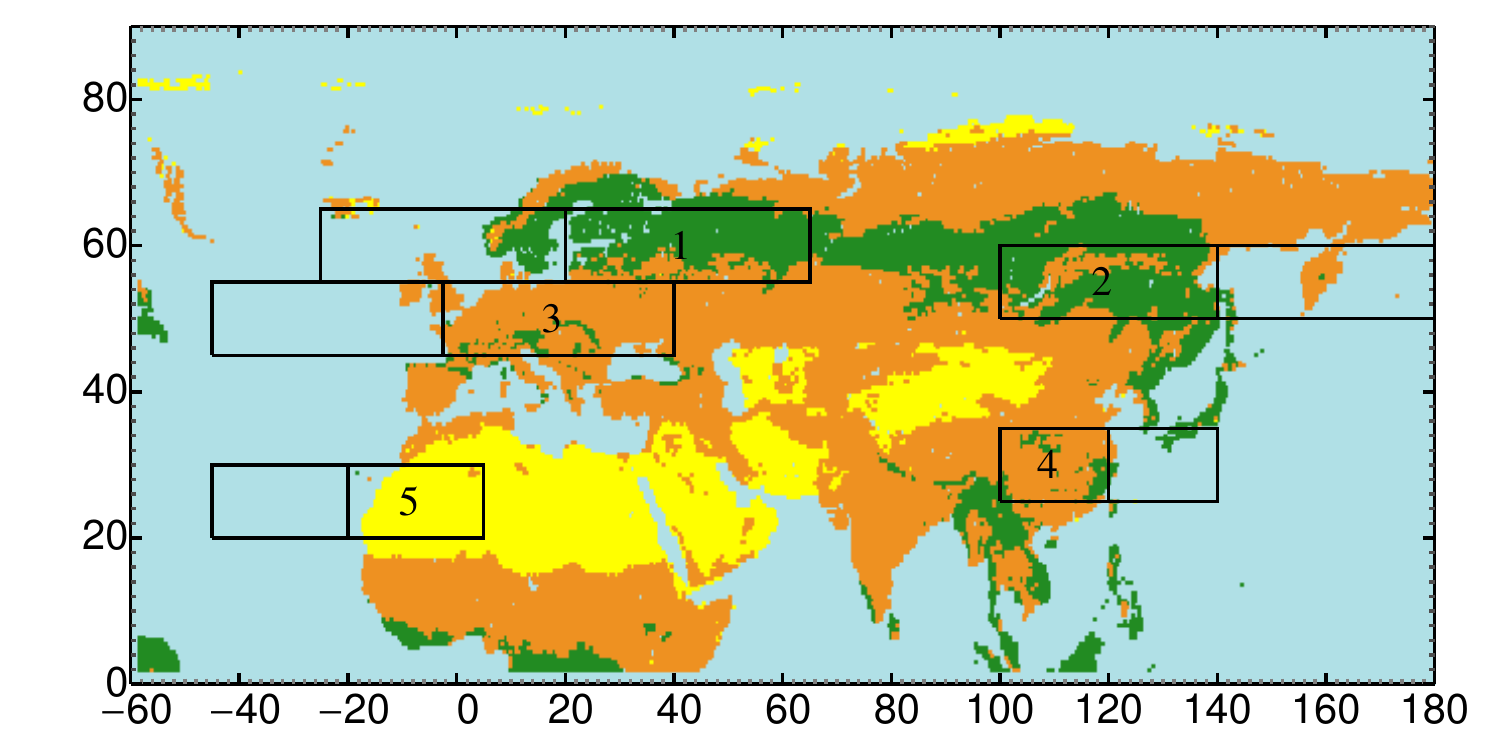}
\end{minipage}
\caption{Regions where the land-ocean temperature contrasts $\Delta T$ were investigated: 1 Boreal Atlantic, 
2 Boreal Pacific, 3 Western Europe,  4 China, 5 Sahara. Different types of vegetation cover (with yellow indicating
unvegetated areas, green -- forested areas, and brown -- areas with other vegetation) are shown following \citep{friedl10} as explained in \citep{mgl13}. }
\label{fig2}
\end{figure*}

\section{Longitudinal land-ocean temperature contrasts in Eurasia}
\label{exp}

We compared temperature differences between land and ocean in four regions of Eurasia with pronounced seasonal dynamics of land-ocean temperature contrasts (Fig.~\ref{fig2}). We also considered the Sahara for comparison with an arid region. The oceanic and terrestrial parts of the regions were chosen to be of equivalent size and latitude.

We used NCEP Reanalysis Derived data concerning the monthly long term means of air temperature, relative humidity 
and zonal and meridional wind at the surface and  geopotential height and air temperature at 13 pressure levels (from 1000 to 70 hPa), as provided by the NOAA/OAR/ESRL PSD, Boulder, Colorado, USA from their website at 
\url{https://psl.noaa.gov/data/gridded/data.ncep.reanalysis.derived.html} \citep{kalnay96}.

\begin{figure*}[tbp]
\begin{minipage}[p]{1\textwidth}
\centering
\includegraphics[width=0.9\textwidth,angle=0,clip]{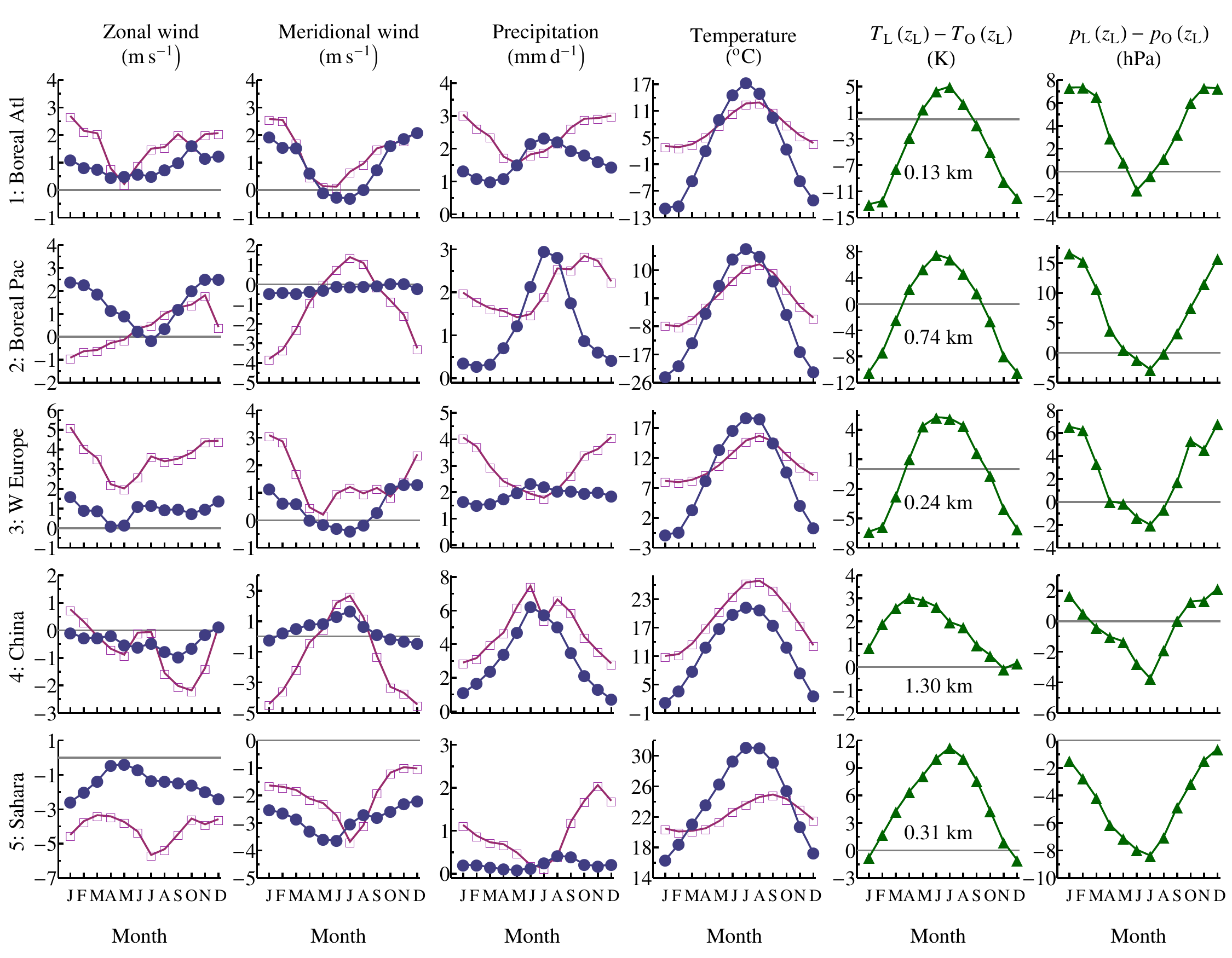}
\end{minipage}
\caption{Zonal and meridional wind, mean monthly precipitation and surface temperatures on land (circles) and over the ocean (squares) in the studied regions, the temperature difference between land and ocean at the mean land elevation $z_{\rm L}$ (shown in the fifth column) and the mean pressure difference at the same altitude. Note that $T_{\rm L}(z_{\rm L})-T_{\rm O}(z_{\rm L})$ is not equal to the difference in surface temperatures in the fourth column.}
\label{fig3}
\end{figure*}

\begin{figure*}[tbp]
\begin{minipage}[p]{1\textwidth}
\centering
\includegraphics[width=0.9\textwidth,angle=0,clip]{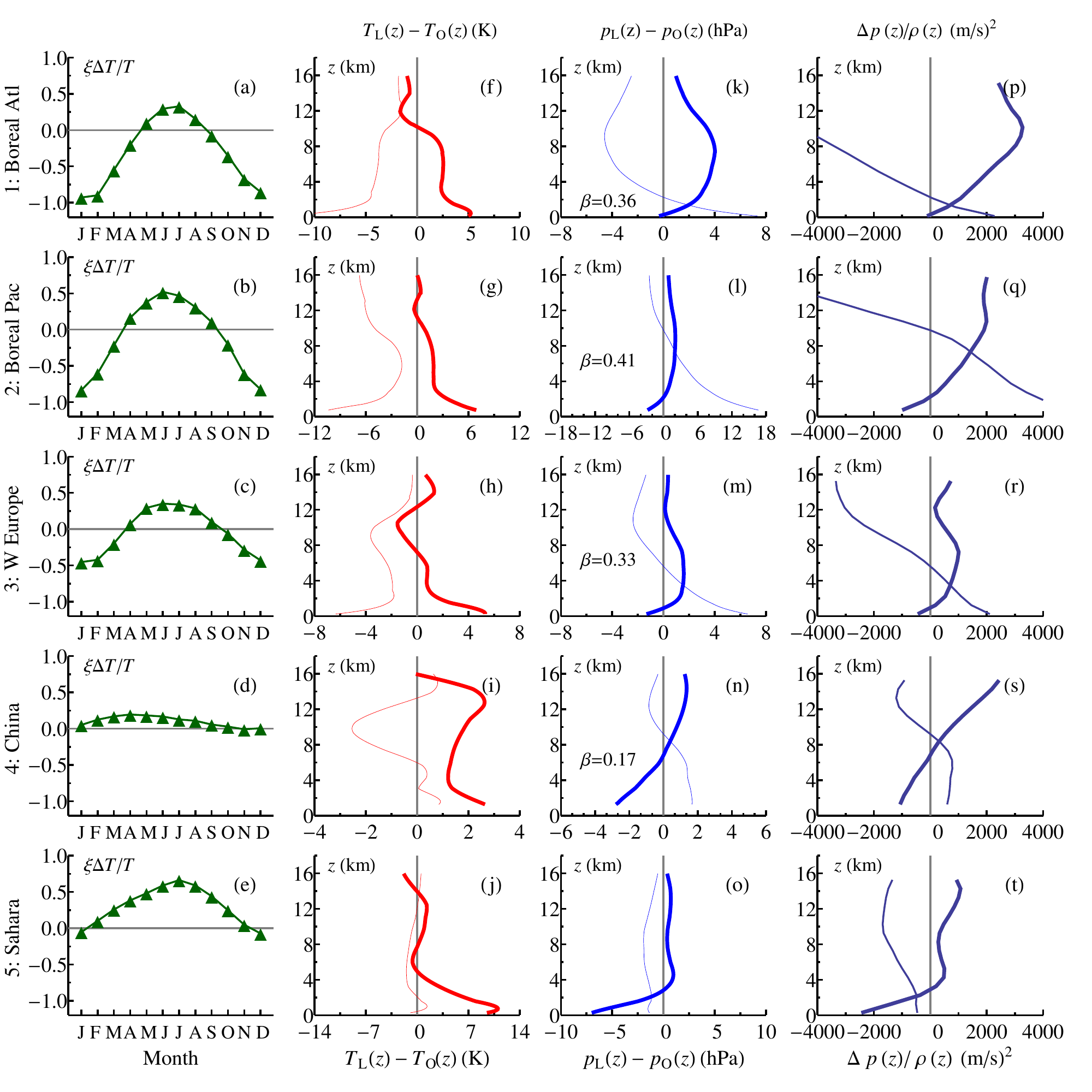}
\end{minipage}
\caption{Parameters of Eqs.~(\ref{beta}) and (\ref{betacalc})-(\ref{betacalc3}). First column: seasonality of the temperature change term in Eq.~(\ref{beta}), with $\Delta T/T  = (T_{\rm L}(z_{\rm L})-T_{\rm O}(z_{\rm L}))/T_{\rm O}(z_{\rm L})$ and $\xi = L/(R T_{\rm O}(z_{\rm L}))$. Second and third columns:  vertical profile of the mean temperature differences between air over land and over the ocean during the month with maximum rainfall (thick solid curve) and January (thin curve). In the third column, solution $p_{\rm A}(\beta, z) - p_{\rm D}(z)$~(hPa) of Eqs.~(\ref{betacalc})-(\ref{betacalc3}) is shown with a dashed curve with the corresponding $\beta$ indicated in the graph. Fourth column: kinetic energy corresponding to the horizontal pressure differences as dependent on altitude $z$ for the month with maximum rainfall  (thick curve) and January (thin curve) (see Discussion).}
\label{fig4}
\end{figure*}

For each month, we averaged temperatures and geopotential heights at the 13 pressure levels separately over land and over the ocean and, by interpolation, obtained vertical profiles of mean temperature on land $T_{\rm L}(z)$ and over the ocean $T_{\rm O}(z)$. From these profiles we calculated the temperature difference between the air over land and the air over the ocean at the mean height $z_{\rm L}$ of land surface above the sea level ($z_{\rm L}$ shown for each region in the fifth column in Fig.~\ref{fig3}).

The maximum of precipitation occurs in July in the two boreal regions, in June in Western Europe and China, and in August in Sahara (Fig.~\ref{fig3}, third column). These rainfall maxima on land in the first three regions are associated with rainfall minima over the ocean, which indicates ocean-to-land moisture transport. The temperature differences between land and ocean are maximum during the summer months  (Fig.~\ref{fig3}, fifth column) except in China. The China region has the highest elevation ($z_{\rm L} = 1.3$~km) among the four regions.  Its land surface is always colder than the oceanic surface, but it is warmer than the atmosphere over the ocean at equivalent elevation (Fig.~\ref{fig3}, fourth vs fifth column). The meridional wind in China increases during maximum rainfall both over land and over the ocean indicating moisture transport from southern regions rather than from the same latitude.

As Fig.~\ref{fig3} indicates, the two multipliers in Eq.~(\ref{beta}) have an opposite seasonal trend. For a given relative humidity, water vapor partial pressure $p_{v}$ at $z_{\rm L}$ grows with increasing temperature and reaches a maximum during the summer months; on land (but not over the ocean) it is maximized simultaneously  with rainfall. But the temperature difference between land and ocean also increases and reaches a maximum in summer, so for a given $\beta$, the term in square brackets in Eq.~(\ref{beta}) declines during the warm season. The maximum magnitude of $\xi \Delta T/T$ ranges from $0.2$ in China to
$0.8$ in Sahara, being on average several times smaller than unity (Fig.~\ref{fig4}a-e). Since $\beta$ is also less than unity, this suggests that an increase in $\Delta T$ above the long-term mean monthly averages that could turn the term in square brackets in Eq.~(\ref{beta}) to zero, is realistic.

We estimated $\beta$ in Eq.~(\ref{beta}) for the month with maximum rainfall from the condition that the pressure difference between two hydrostatic air columns with the vertical temperatures profiles $T_{\rm L}(z)$ and $T_{\rm O}(z)$ (Fig.~\ref{fig4}f-j) and the pressure difference at $z_{\rm L}$ equal to $\Delta p(\beta)$, turns to zero at height $z_0$ where the ratio of local $p_v(z_0)$  to surface $p_v(z_{\rm L})$ equals $\beta$:
\begin{equation}\label{betacalc}
p_{\rm A}(\beta,z_0)-p_{\rm D}(z_0)=0,\quad \beta = \frac{p_v(z_0)}{p_v(z_{\rm L})},
\end{equation}
where
\begin{gather}\label{betacalc2}
p_{\rm A}(\beta,z)\equiv [p_{\rm O}(z_{\rm L}) + \Delta p(\beta)]\exp[-(z-z_{\rm L})/h_{\rm L}],\quad h_{\rm L} \equiv \frac{RT_{\rm L}(z)}{M g},\\ \label{betacalc3}
p_{\rm D}(z)\equiv p_{\rm O}(z_{\rm L}) \exp[-(z-z_{\rm L})/h_{\rm O}],\quad h_{\rm O} \equiv \frac{RT_{\rm O}(z)}{M g}.
\end{gather}

The ratio of water vapor partial pressures at $z_0$ and $z_{\rm L}$ in Eq.~(\ref{betacalc}) was calculated from the Clausius-Clapeyron equation (\ref{CC}) assuming that at $z_0$ water vapor is saturated and at $z_{\rm L}$ the relative humidity is equal to the mean monthly relative humidity. The solutions are shown in Fig.~\ref{fig4}k-o together with the actual pressure difference profiles. These solutions correspond to $\beta$ values ranging from $0.17$ to $0.41$ and $z_0$ below 3~km. Similar values are obtained for the June and August temperature profiles (data not shown). These findings are consistent with the observation that most kinetic power in the extratropical atmosphere is generated below the $800$~hPa level \citep[e.g.,][Fig.~A1c,d]{tellus17}.

Using the estimated $\Delta T \equiv T_{\rm L}(z_{\rm L}) - T_{\rm O}(z_{\rm L})$ for the month with maximum rainfall in each region, we can find additional temperature difference $\delta T$ that would turn $\Delta p$ in Eq.~(\ref{beta}) to zero and block atmospheric moisture transport to the region: 
\begin{equation}\label{betafin}
\delta T(\beta) \equiv \frac{\beta}{1 - \beta} \frac{T}{\xi} - \Delta T. 
\end{equation}

\begin{figure*}[tbp]
\begin{minipage}[p]{1\textwidth}
\centering
\includegraphics[width=0.8\textwidth,angle=0,clip]{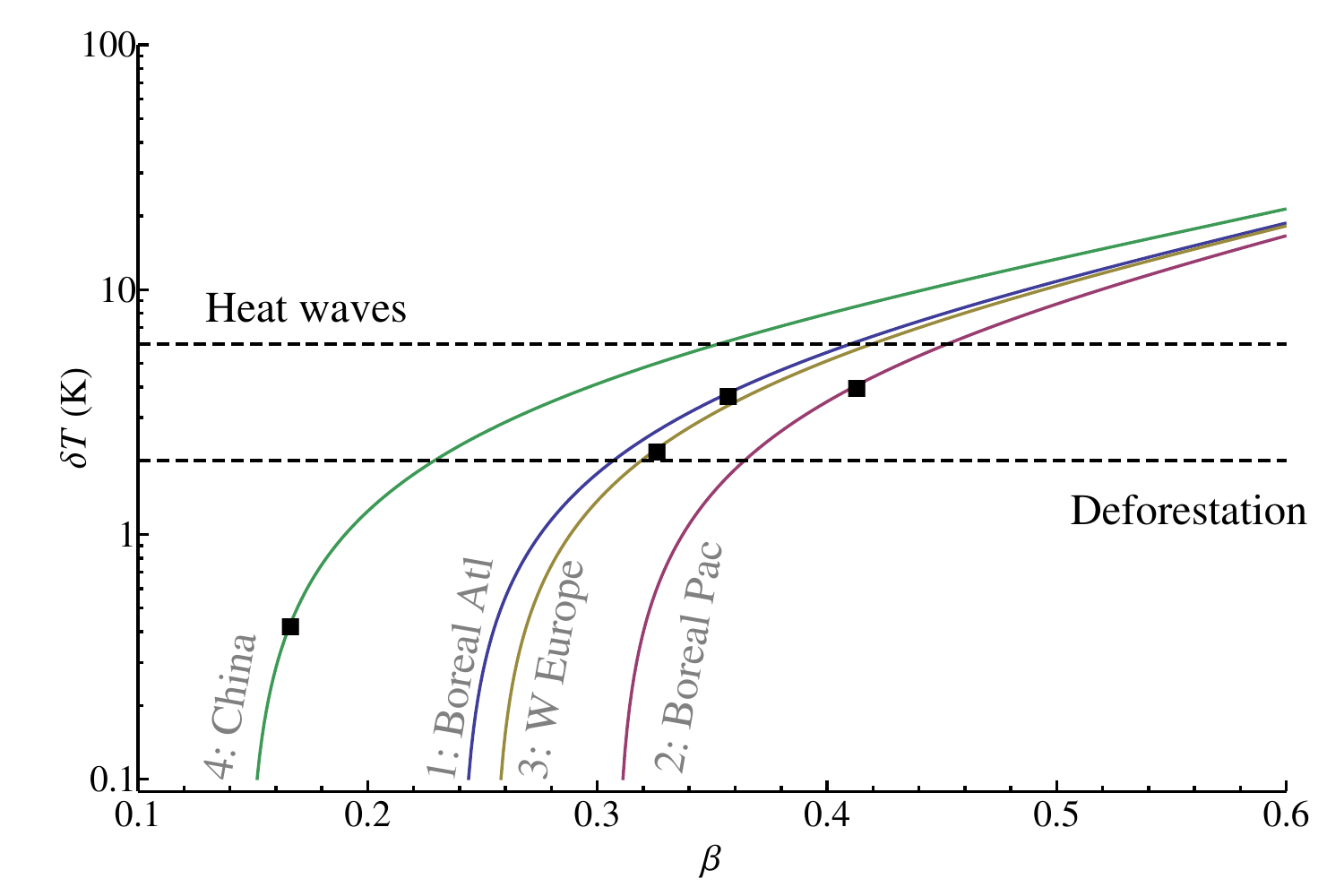}
\end{minipage}
\caption{Solutions of Eq.~(\ref{betafin}) for the four Eurasian regions: $\delta T$ is the additional temperature difference
between land and ocean at which the condensation-induced moisture transport is expected to discontinue. The two dashed lines
indicate a temperature anomaly ($6$~K) characteristic of recent heat waves in the Northern Hemisphere \citep{philip2021}, and the temperature difference ($\sim2$~K) caused by (partial) cutting of native forests \citep{baker2019,alkama16}. Solid squares indicate the $\beta$ values from Fig.~\ref{fig4}k-n. The black square for Western Europe located at the {\textquotedblleft}Deforestation{\textquotedblright} line indicates that at the estimated $\beta = 0.33$ in June the two degrees of extra warming due to forest cutting can potentially prevent regular moisture transport.}
\label{fig5}
\end{figure*}

These solutions are shown in Fig.~\ref{fig5}. One can see that at the observed $\beta$ values a few degrees extra warming of land can prevent circulation. For example, for Western Europe an extra land warming of two degrees is sufficient.
Extra warming on land is associated with heat waves and blocking anticyclones \citep[e.g.,][]{chan2019}.
During heat waves the temperature anomaly may reach six degrees Celsius \citep[e.g.,][]{philip2021}.
Disturbance of the forest cover by deforestation increases summer temperatures by up to two degrees \citep{baker2019,alkama16}.
We could have found that the values of $\beta$ and $\Delta T$ in Eurasia are, respectively, too large and too small, such that
the observed temperature anomalies $\delta T$ have a negligible impact on the parameters of the condensation-induced moisture
transport. Instead, we found that these values are such that a disruption of the water cycle by extra warming is plausible and indeed might be responsible for the increasing frequency of extreme events. It is a prediction to be tested.

\section{Discussion}

The continuity equation for a gaseous atmosphere yields an estimate on wind power provided  the expression for condensation rate is known. We have demonstrated that this estimate is sensitive to  condensation rate with minor changes in condensation rate resulting in marked changes in wind power (Table~\ref{tab}). These results are pertinent to the challenge of predicting regional changes of the terrestrial water cycle, especially in monsoonal climates where models currently disagree even on the sign of possible changes under global warming \citep[e.g.,][]{hill2019}.  Recognizing this sensitivity can guide and improve model parameterizations.

We have further shown that that within the framework of condensation-induced atmospheric dynamics wind power is reduced when horizontal air motion occurs in the direction of higher temperatures. Addition of water vapor by evaporation in the horizontal flow partially compensates for the pressure drop by condensation in the ascending air. This reduces the release of potential energy to power winds. The generation of wind power (the scalar product of wind velocity and pressure gradient) is what sustains wind despite energy lost to friction. Once the temperature gradient becomes sufficiently large, the condensation-induced wind power  becomes zero and the related air flow ceases.

This effect is not accounted for outside CIAD. Consideration of possible changes in the ocean-to-land circulation (Fig.~\ref{fig1}) is made in terms of the temperature effects \citep[the {\textquotedblleft}breeze-like{\textquotedblright} circulation][]{hill2019}.  In a hydrostatic atmosphere, other things being equal, a higher surface temperature creates a pressure surplus in the upper atmosphere that pushes the air away from the warmer air column. This air outflow reduces total air mass and produces a pressure shortage at the surface that causes the low-level air convergence towards the warmer area.  However, in this consideration the magnitude of the resulting air inflow and whether the warming is efficient enough to explain the observed convergence, cannot be unambiguously quantified \citep[e.g.,][]{lindzen87}. The conventional qualitative picture does not rule out other mechanisms.

CIAD provides a distinct mechanism to reduce surface pressure and to power low-level moisture convergence.  But CIAD does not specify a mechanism for the upper-level outflow.  Figure~\ref{fig4}p-t  shows that the kinetic energy required to move against the upper tropospheric temperature-related pressure differences are in the order of $10^3$~m$^2$~s$^{-2}$. Such energies are present in tropical cyclones but not in large-scale transcontinental air circulation. For the CIAD to be realized some mechanism for the outflow should be present: either the differential warming, or latent heat release, or, as in hurricanes, the cyclostrophic imbalance. This results in condensation being often concentrated in the warmer regions  (unless, like in Ferrel cells, there is an external dynamical driver to push the  upper air against the pressure surplus in the upper atmosphere \citep[][]{tellus17}).

This coupling of CIAD with conventional mechanisms (like a higher moisture inflow towards a warmer land surface) masks the CIAD presence in the conventional qualitative picture and could be the reason for CIAD having been neglected. The fact that CIAD produces realistic quantitative estimates of wind power is an indication that without CIAD the same outflow mechanisms would have generated weaker circulations.

The differential warming of land versus the ocean and the preferential release of latern heat either over land or over the ocean have different implications for the resulting circulations. It has been recognized, since the works of Charney \citep{charney75}, that a warm land surface does not necessarily initiate a moisture inflow if the land is also dry: the surface warming will be negated by cooling of the adiabatically ascending dry air (Fig.~\ref{fig1}a). However, if the land is both warm and moist, the conditions for ocean-to-land moisture inflow are conventionally considered to be favorable. We show that this may not always be the case. Too high land-ocean temperature contrasts can  inhibit, and block,  moisture inflows and the ascending motion of moist air.

This can be a mechanism contributing to the formation of blocking anticyclones and heat waves, for which, as commonly recognized, there is no comprehensive dynamic theory or understanding \citep{woollings2018,miralles2019}. One of the conceptual questions is the following: if the air rises where it is warm, why does it descend where it is the warmest, i.e., during heat waves associated with blocking anticyclones? Here the above-described difference between CIAD and temperature-driven circulation provides a clue.

For the CIAD circulation to work, there must be a pressure drop from condensation. This happens when the moist air rises and vapor condenses. As the moist air moves horizontally towards the area of ascent, if the surface is moist and its temperature increases along the streamline, the air will acquire water vapor. If this temperature rise is too high, the amount of acquired water vapor due to evaporation can exceed the amount lost due to condensation. The net pressure difference will be zero, and the CIAD circulation will stop  or reverse. \citep[Indeed, heat waves are accompanied by a spike in evaporation, see, e.g.,][Fig.~2]{sitnov2014,miralles2019}. The prevalence of CIAD mechanisms over the temperature-driven motions will then account for the persistence of the descending air motion in the warmest area.

We have shown that the long-term temperature contrasts during the warm season in Eurasia are close to the threshold where the moisture transport ceases and can be driven beyond the threshold by temperature anomalies characteristic of recent heat waves registered in the Northern Hemisphere (Fig.~\ref{fig4}).  We have discussed that in this situation the surface cooling provided by transpiring vegetation can be crucial to avoid a tipping point where the climate switches to arid. Natural, undisturbed forests provide the strongest buffer against surface warming \citep{alkama16,baker2019}. 

Schematically, removing forests can initiate a dangerous feedback: as the horizontal temperature contrasts grow, moisture transport declines which further increases the temperature surplus. If the descending air in a blocking anticyclone has created a critical temperature surplus, in the absence of vegetation nothing can disrupt the resulting circulation. Conversely,  a water-sufficient sustainable forest can forcefully cool the area through transpiration and thus reduce the excess heat. A large-scale example of this process in action is provided by the Amazon rainforest that promotes the onset of the wet season on by enhancing transpiration during the dry season \citep{wright17}.

It is critically important to continue theoretical investigations of moist atmospheric dynamics and the role of vegetation,
considering jointly the biotic pump mechanism and the temperature-driven effects.  Previously, we have indicated that a major role of vegetation in the atmospheric moisture transport is to keep the atmosphere moist via transpiration \citep{hess07,jhm14}. Here we highlight an additional role: to buffer the land-ocean temperature contrasts.  Recognizing these physically distinct effects of temperature on air circulation can help better understand and project the diverse impacts of land cover change on the local and regional water cycle \citep{lawrence15,tewierik2021,caballero2022}. The effects of the vegetation cover change on the regional water cycle can be exacerbated by atmospheric teleconnections \citep[e.g., the Rodwell-Hoskins mechanism,][]{rodwell96}.

In a broader context, the current international focus on mitigating carbon emissions raises the importance of renewable energy sources and causes an increased pressure on global forest ecosystems \citep{jonsson17,lauri17}.  Deforestation emits carbon dioxide warming the planet but in the boreal region is considered to cool the planet via an increase in albedo,  the overall effect is judged to be small despite the uncertainties \citep{ipcc19}.  This narrative has allowed the on-going extirpation of native boreal forests to proceed with little international concern.

At the same time, it has become clear that vast forests play an active role in transcontinental moisture transport and in controlling regional temperature and water regimes \citep[e.g.,][]{pnobre09,ent10,Pielke11,alkama16,Mahmood16,leitefilho21,meier2021}.  Russia, for example, is home to some of the world{\textquoteright}s most extensive natural forest ecosystems \citep{potapov08}. The pristine forest ecosystems  are characterized by resilience to perturbations like fires, windfall or pests \citep{sukachev75,gromtsev02,rich07,shorohova08,debkov19}.  They also stabilize regional and global climates \citep{vg95,funk19,makarieva20b}.

Recent research has highlighted how forests buffer downwind regions against fluctuations in precipitation \citep{oconnor2021}. Conversely, loss of native forest cover should result in continent-scale destabilization of the water cycle and temperature regime. Indeed, simultaneously with pristine forests being lost in Russia \citep{potapov17}, the Eurasian continent is drying and increasingly suffering violent winds, floods and droughts \citep{gu19,krause20,cornwall2021}.  Droughts and other disruptions of the water cycle compromise the forest capacity to absorb carbon \citep{sheil19} and thus add to the global climate destabilization associated with carbon emissions. There is a long-term legacy from  past land-use practices that determines ecosystem{\textquoteright}s response to current climate \citep{aleinikov19,buras20}. It is therefore crucial to differentiate ecosystems capable of self-recovery from those on the degradation trajectory. Once disrupted, given the large time scale of forest successional recovery towards the natural state, the moisture-regulating  functions of intact forests cannot be rapidly restored. Given the complex vegetation properties and the processes that  prevent abrupt landscape transitions from wetness to aridity (Fig.~\ref{fig1}), a random replacement of intact forests by artificial plantations is not likely to recover the water cycle stability. This may explain mixed success of large-scale afforestation/rewetting efforts in China \citep{jiang13,ahrends17,zinda19,zhao2021}.

A strategy to mitigate climate change and stabilize the continental water cycle must include a focused research-policy program aimed at protecting natural forests (in Russia, Canada  and beyond).  As moisture transport ignores political borders making downwind countries highly dependent on upwind vegetation cover  \citep{ent10}, forest conservation and restoration policies in one country (e.g., China) will  not be successful if they are accompanied by increased pressure on intact ecosystems in another (e.g., Russia).  While the appreciation of the importance of primary ecosystems is now on the rise \citep{easac17,jonsson20,sabatini20}, the understanding, and corresponding research, of their active participation in the many aspects of climate stabilization, as well as of the potential of {\it proforestation} \citep{moomaw2019} for climate change mitigation, remain inadequate. Large-scale drought-mitigation measures can only be successful within a broader strategic framework that recognizes the role of forest cover, and pristine forests in particular, in the water cycle and atmospheric dynamics. Elaborating such a framework requires a major interdisciplinary effort.

\section*{Acknowledgments}
Work of A.M. Makarieva is partially funded by the Federal Ministry of Education and Research (BMBF) and the Free State of Bavaria under the Excellence Strategy of the Federal Government and the L\"ander, as well as by the Technical University of Munich -- Institute for Advanced Study.

%\newpage
%\appendix
\setcounter{section}{0}%
\setcounter{equation}{0}%
\renewcommand{\theequation}{A.\arabic{equation}}%

\section*{\begin{center}  Appendix A:  Deriving Eqs.~(\ref{cJ}) and (\ref{tg})  \end{center}}
\label{der}

For the convenience of our readers here we repeat the derivations of  \citet{mgn14}. In the stationary case, the continuity equations for the water vapor and the dry air constituents have the form 
\begin{subequations}\label{eq67}
\begin{align}
\nabla \cdot (\textbf{v}N_v) \equiv N_v (\nabla \cdot \textbf{v})+ (\textbf{v} \cdot \nabla) N_v =  \mathcal{S},   
\label{eq6}\\
\nabla \cdot (\textbf{v}N_d) \equiv N_d (\nabla \cdot \textbf{v})+(\textbf{v} \cdot \nabla) N_d = 0,  \label{eq7}
\end{align}
\end{subequations}
where $N_v$, $N_d$ and $N = N_v + N_d$ are the molar densities of water vapor, dry air constituents and moist air as a whole, respectively. Air velocity $\textbf{v} = \textbf{u} + \textbf{w}$ is equal to the sum of the horizontal $\textbf{u}$ and vertical $\textbf{w}$ velocity components. The quantity $\mathcal{S}$ (mol~m$^{-3}$~s$^{-1}$) represents the volume-specific rate at which molar density $N_v$ of water vapor is changed by phase transitions. By multiplying \eqref{eq7} by $\gamma_d \equiv  N_v / N_d$  and excluding $N_v( \nabla \cdot \textbf{v})$ from \eqref{eq6}, we obtain
\begin{equation}\label{eq8}
(\textbf{v} \cdot \nabla) N_v - \gamma_d (\textbf{v} \cdot \nabla) N_d=  \mathcal{S}.
\end{equation}
Using the ideal gas equation of state 
\begin{equation}\label{eq9}
p= N R T , \quad p_v = N_v RT  , \quad p_d= N_d RT   ,
\end{equation}
one can replace molar densities $N_i$ in \eqref{eq8} with partial pressures $p_i$ ($i=v,d$) and the rate of phase transitions $\mathcal{S}$ with the power of phase transitions $\sigma \equiv \mathcal{S}RT$ (W~m$^{-3}$):
\begin{equation}\label{eq10}
(\textbf{v}\cdot \nabla) p_v - \gamma_d (\textbf{v}\cdot \nabla) p_d=  \sigma .
\end{equation}
Owing to the universality of the gas constant $R$ the contribution due to temperature gradient $\nabla T$ cancels. 

Substituting \eqref{cp} into \eqref{eq10} and taking into account the identity
\begin{equation}\label{eq12}
\nabla p_v -  \gamma_d \nabla p_d \equiv (1 + \gamma_d) ( \nabla p_v - \gamma \nabla p),
\end{equation}
where $\gamma  \equiv  p_v/p \equiv  \gamma_d/(1+ \gamma_d)$, we obtain the following relation for \eqref{eq10}:
\begin{equation}\nonumber
(\textbf{w}  + \textbf{u}) \cdot (\nabla p_v -  \gamma \nabla p)=
\frac{1}{1+ \gamma_d} (\textbf{u} \cdot \nabla) p  .
\end{equation}
By transferring $\gamma (\textbf{u} \cdot \nabla) p$ to the right-hand side of the last relation and
taking into account relation between $\gamma$  and  $\gamma_d$, we obtain:
\begin{equation}\label{eq13}
p (\textbf{w}\cdot \nabla) \gamma + (\textbf{u} \cdot \nabla) p_v = (\textbf{u}\cdot  \nabla) p , \quad
p\nabla \gamma \equiv \nabla p_v - \gamma \nabla p ,
\end{equation}
which coincides with Eq.~\eqref{tg}.

%%%%%%%%%%%%%%%%%%%%%%%%%%%%%%

%\bibliographystyle{copernicus}
%\bibliography{met-refs}

\end{document}